\documentclass[superscriptaddress,showpacs,nofootinbib,preprint]{revtex4}
\usepackage{amssymb,graphicx,dcolumn,bm,textcomp}

\begin{document}

\newcommand*{\pku}{School of Physics and State Key Laboratory
of Nuclear Physics and Technology, \\Peking University, Beijing
100871, China}\affiliation{\pku}

\newcommand*{\CHEP}{Center for High Energy
Physics, Peking University, Beijing 100871,
China}\affiliation{\CHEP}

\newcommand*{\CHPS}{Center for History and Philosophy of Science,
Peking University, Beijing 100871, China}\affiliation{\CHPS}

\title{Note on a New Fundamental Length Scale $l$ Instead of the Newtonian Constant $G$\footnote{To be published in
    SCIENCE CHINA {\it Physics, Mechanics \& Astronomy} (Science in China Series G).}}

\author{Lijing Shao}\affiliation{\pku}
\author{Bo-Qiang Ma}\email[Corresponding author.
Electronic
address:~]{mabq@pku.edu.cn}\affiliation{\pku}\affiliation{\CHEP}\affiliation{\CHPS}

\begin{abstract}
The newly proposed entropic gravity suggests gravity as an emergent
force rather than a fundamental one. In this approach, the Newtonian
constant $G$ does not play a fundamental role any more, and a new
fundamental constant is required to replace its position. This
request also arises from some philosophical considerations to
contemplate the physical foundations for the unification of
theories. We here consider the suggestion to derive $G$ from more
fundamental quantities in the presence of a new fundamental length
scale $l$, which is suspected to originate from the structure of
quantum space-time, and can be measured directly from
Lorentz-violating observations. Our results are relevant to the
fundamental understanding of physics, and more practically, of
natural units, as well as explanations of experimental constraints
in searching for Lorentz violation.
\end{abstract}


\maketitle

The unification of gravitational force with the other three forces
in the standard model of particle physics is a most intriguing aim
of the physicist community. However, explorations in this direction
have been confronted with severe difficulties, and we are still
without a satisfactory solution. There have been other attempts to
understand gravity as an emergent phenomenon based on quantum theory
and thermodynamics~\cite{Padma,Padma2}. Recently, Verlinde proposed
an appealing argument to explain gravity as an entropic
force~\cite{v10} from holographic
perspectives~\cite{holography,susskind}. In Verlinde's approach,
gravity comes from the tendency of maximizing entropy in black hole
physics, \textit{i.e.}, gravity is actually an entropic force
instead of a fundamental one. He conjectured that the entropy change
$\Delta S$ associated with a displacement, $\Delta x$, towards a
holographic screen by a particle with mass $m$, is
\begin{equation}\label{deltaS}
\Delta S = 2 \pi k_B \frac{mc}{\hbar} \Delta x,
\end{equation}
where $c$, $k_B$, and $\hbar$ are the speed of light in vacuum, the
Boltzmann constant, and the reduced Planck constant, respectively.
By utilizing holographic principles and the Unruh
temperature~\cite{u76}, Verlinde succeeded in reproducing Newton's
second law. By introducing a number of used bits on the holographic
screen together with the equipartition Ansatz of mass $M$ within the
holographic screen, he also derived Newton's law of gravity.

From a logical viewpoint, if the gravitational force represents an
emergent force  rather than a fundamental one, the Newtonian
constant $G$ does not play a fundamental role any more, but instead,
should be considered as a deduced quantity from more basic
constants. Conventionally, the gravitational constant $G$ is one of
the five ``God-given'' ingredients, \textit{i.e.}, $c$, $G$,
$\hbar$, $k_B$, and $1/4\pi \epsilon_0$ (where $\epsilon_0$ is the
permittivity of free space), to construct the Planck
units~\cite{p99}. In 1899, Planck set the above five constants as
bases, and elegantly simplified recurring algebraic expressions in
physics. The nontrivial non-dimensionalization has conceptually
profound significance for theoretical physics. There are a number of
basic quantities in this unit system, such as the Planck length
$l_{\rm P} \equiv \sqrt{G\hbar/c^3} \simeq 1.6 \times 10^{-35}$~m,
the Planck time $t_{\rm P} \equiv \sqrt{G\hbar/c^5} \simeq 5.4
\times 10^{-44}$~s, the Planck energy $E_{\rm P} \equiv \sqrt{\hbar
c^5/G} \simeq 2.0 \times 10^{9}$~J, and the Planck temperature
$T_{\rm P} \equiv \sqrt{\hbar c^5/Gk_B^2} \simeq 1.4 \times
10^{32}$~K~\cite{p99}.

Now if $G$ is removed from the five basic ingredients, we need
another fundamental quantity to complete the unit system. It can be
an energy scale, or a length scale, or a time scale, and such a
fundamental quantity is necessary to retrieve $G$ as a deduced
quantity. We here choose a new fundamental length scale $l$,
following a recent suggestion~\cite{k07,k10}. If $l$ happens to
coincide with $l_{\rm P}$, then the unit system remains intact by
replacing $G$ with $l_{\rm P}$. Otherwise, there will be conceptual
consequences on the unit system.

On the other hand, there have been many debates on the issue of
fundamental units. There is a famous trialogue on the number of
dimensionful units of fundamental significance among Duff, Okun, and
Veneziano~\cite{dov02}. Okun insisted that space, time, and mass are
three basic physical dimensions which can arise from special
relativity ($c$), quantum mechanics ($\hbar$), and gravity ($G$).
Among them, ``The position of $G$ is not as firm as that of $c$ and
$\hbar$'', stated by Okun~\cite{dov02}. $G$, $\hbar$, and $c$ form a
``cube of theories'', with vertices denoting our understanding of
physics related to distinguished physical realities~\cite{cube}.
Veneziano argued that two fundamental units of space and time, the
light speed $c$ and a string length scale $\lambda_{\rm s}$, are
both necessary and sufficient to be fundamental in the framework of
quantum string theory and/or M-theory. However, Duff advocated that
all dimensionful units are merely conversion factors, and only
dimensionless quantities really matter to physical processes. It
seems that the position of $G$ as fundamental is reasonably doubted
by all of the three sides in the trialogue.

It is necessary to mention that, the well-known, widely-used, and
globally-adopted international system of units (abbreviated as
``SI'' from the French ``Syst$\grave{\rm e}$me international
d{\textquotesingle}unit$\acute{\rm e}$s''), which possesses seven
units, is not relevant to discussions here, because that they are
actually related to a system of units of measurements, rather than
fundamental physics.

Bearing the suspicion of gravity and $G$ as fundamental in mind, let
us analyze how $G$ is introduced by Verlinde. For a region of mass
$M$ enclosed by a spherical surface $A=4\pi R^2$, all information
stored on $A$ is assumed to be given by the number of
bits~\cite{v10}
\begin{equation}\label{n}
N = \frac{Ac^3}{G\hbar} \equiv \frac{A}{S_{\rm P}},
\end{equation}
where $G$, or equivalently the Planck area $S_{\rm P} \equiv
l_{P}^2$~\cite{hm10}, is introduced as a parameter 
in the information assignment process. By adopting the energy
equipartition principle, $E=N \cdot k_BT/2 = Mc^2$, and the
well-known thermal relation, $F \Delta x = T \Delta S$,
Eqs.~(\ref{deltaS}) and (\ref{n}) lead to Newton's gravity law,
\begin{equation}
F = G \frac{Mm}{R^2}.
\end{equation}
Hence $G$ is identified as the gravitational constant~\cite{v10}.
From above, we see that the Planck area
 $S_{\rm P} \equiv
l_{P}^2$, or the Planck length  $l_{P}$, might be considered as a
more basic quantity to replace $G$ in the entropic gravity
approach~\cite{hm10}. However, here the holographic bits assignment
process is involved, whose detailed dynamics is yet unknown. Hence
$l_{\rm P}$ as the fundamental length scale is not decisive;
actually, it is quite questionable. The String Theory Committee
might argue that the string length $\lambda_{\rm s}$ is much more
fundamental. Ref.~\cite{kv05} hypothesized that gauge symmetry, as
well as bosons, can be emergent, and suggested the ultraviolet
cutoff of fermions, which might be associated with Lorentz symmetry
violation, to be basic. From viewpoints of fermionic vacuum
polarization, the cutoff can be as large as $10^{25} E_{\rm P}$ in
the presence of three families of fermions, and $10^8 E_{\rm P}$
with five families~\cite{kv05}. Besides, some branches of gravity
society themselves are also endeavoring in the search for a more
basic origin of gravity, instead of accepting it as given.
Therefore, there are strong motivations for a new quantity to
replace $G$ as fundamental.

We here follow the suggestion to introduce a fundamental length
scale $l$, which, from dimensional analysis, is proportional to
$l_{\rm P}$, \textit{i.e.}, $l = l_{\rm P}/\eta$~\cite{k07,k10}.
Then from Eq.~(\ref{n}), we have
\begin{equation}\label{n2}
N = \frac{A}{\eta^2l^2}.
\end{equation}
We see that $\eta^2$ reflects the relation of the physically
fundamental area, $l^2$, and the unit area needed to store one bit
of information from holographic perspectives.

Consequently, the conventional Newtonian constant $G$ appears to
be~\cite{k07,k10}
\begin{equation}\label{g}
G = \frac{\eta^2l^2c^3}{\hbar},
\end{equation}
where it is considered as derived from fundamental constants $c$,
$\hbar$, and $l$. Therefore, $G$ does not play a fundamental role as
a priori any more in the entropic gravity scenario~\cite{k10}.

Worthy to note that, as suggested by Verlinde, after including the
Unruh temperature $T = \hbar a / 2\pi c k_B$~\cite{u76}, where $a$
denotes the acceleration of motion, the second law of Newton,
$F=ma$, also emerges directly~\cite{v10,hm10}. Further by
introducing the equivalence principle, Einstein equations are also
obtained~\cite{v10}, in a similar way with an analogy between
general relativity and thermodynamics~\cite{j95} .

Now let us turn to our main concern on the relation between $G$ and
$l$. Naively speaking, $\eta$ is ``defined'' in Eq.~(\ref{n2}) and
reflects the dynamics of information assignment onto the holographic
screen. It plays a role of some unknown principles of storing
information~\cite{k10}. Once the procedure of digital storage onto
holographic screens is well understood, it is practical to calculate
$\eta$ from certain holographic rules. However, nowadays, due to the
imperfect knowledge of the holographic mechanism, we cannot get its
value theoretically, even through evaluation. Fortunately, it can be
settled through Lorentz violating experiments~\cite{k10}, after we
identify it to be related to the minimal structure of quantum
space-time, or equivalently, quantum gravity. The indeterminate
value of $\eta$ influences the fundamental length scale $l$ and the
corresponding energy scale $E$, as well as explanations of
experimental data in the search of Lorentz symmetry violation.

The search for Lorentz violation has energetically lasted for more
than ten years; however, no definite deviation from special
relativity is yet confirmed~\cite{m05,a08}. Hence one begins to
doubt the possibility and reasonableness of Lorentz violation. If
the arguments above turn out to be valid, a new fundamental length
scale $l$ can be introduced; then the conventional ``criterion'',
\textit{i.e.}, the Planck quantities $l_{P}$ and $E_{P}$, should be
modified. In the case of large $\eta$, \textit{e.g.}, $\eta > 10^8$,
most constraints from astrophysical and laboratory observations can
be compatible to Lorentz symmetry violation with linearly modified
terms. As another merit, the Lorentz-violating research has already
validated plenty of experiments and observations to explore the
relevant energy scale. Thus it appears a promising strategy to
determine the fundamental length scale $l$ through Lorentz
violation, \textit{e.g.}, time delays of high energy photons
relative to low energy
ones~\cite{m05,a08,grb,xm09,sxm10,sm10,xsm10,fermi09,sm11prd}, the
reaction patterns of ultra-high energy cosmic rays and TeV
$\gamma$-rays with background low energy photons, and the instances
of some forbidden reactions~\cite{m05,a08,sm10}.

Conventionally, $l_{\rm P}$ represents a fundamental quantity in the
search of quantum gravity. Actually, it is used as a ``criterion''
to justify quantum gravity
phenomenons~\cite{m05,a08,grb,xm09,sxm10,sm10,xsm10,fermi09,sm11prd,a02,a02b}.
For example, in Ref.~\cite{fermi09}, the Fermi GBM/LAT Collaboration
estimated the quantum gravity energy scale from time delays of high
energy photons of a $\gamma$-ray burst, GRB~090510, to be larger
than the Planck energy $E_{\rm P}$. It was argued to rule out the
linear energy dependence of Lorentz-violating modifications on the
speed of light. However, if the referenced energy scale is some
orders larger in magnitude than $E_{\rm P}$, their conclusion would
not be valid. In that case, the experimental implications on whether
to rule out linear quantum-gravitational modifications on the light
speed should be re-explained.

It is clear that $\eta \rightarrow 1$ leads to $l \rightarrow l_{\rm
P}$. However, according to discussions above, $\eta$ is not
protected to be around unit. The possibility that $\eta$ is far away
from unit is not theoretically forbidden from holographic
viewpoints, and other theoretical considerations (see,
\textit{e.g.}, Ref.~\cite{kv05} for a rather large $\eta \sim
10^{25}$ in presence of three families of fermions, and $\eta \sim
10^8$ in presence of five families of fermions). For a generic $\eta
\neq 1$, we have a new length scale $l$ for reference instead of the
conventional $l_{\rm P}$, as well as a new energy scale $E = {\hbar
c / l}$ instead of $E_{\rm P}$. Experimentally, the positions of
$\hbar$ and $c$ are more firm than that of $G$, so we here adopt
them as intact. To do comparisons more conveniently, we write the
new scales in terms of conventional Planck scales,
\begin{equation}\label{le}
l = l_{\rm P}/\eta, \quad\quad E = \eta E_{\rm P}.
\end{equation}
If $\eta=10^8$, then $l = 10^{-8} l_{\rm P}$ and $E = 10^{8} E_{\rm
P}$. This represents an illustrated case where the space needed to
store per piece of information is rather larger than the physically
fundamental area $l^2$, \textit{i.e.}, $10^{16} l^2 = S_{\rm P}$ for
one bit. However, the situation is possible (reasonable) in reality,
maybe ascribing to the poor quality of physical (foamy) areas for
storing holographic information.

Now let us further discuss the physical implications of the newly
introduced $l$. The most likely scenario to accommodate such a
fundamental length scale is the space-time foam conjecture, firstly
suggested by Wheeler in nineteen fifties~\cite{w57}. It is a
speculative extension of the space-time concepts with hypothesis of
the coexistence of the matter-geometry interrelation and the
uncertainty-principle-induced high-energy virtual particles at very
short distances. As the length scale approaches $l$, the classically
continuous description of space-time breaks down, and novel quantum
gravity theories are needed to complete a consistent description of
physical processes. Here, the proposed $l$ can be the physically
smallest scale of space-time structure. For a concrete example, $l$
can be explained as the scale where the space-time coordinates fail
to commute, as in the non-commutative quantum field
theories~\cite{dn01}.

However, another puzzle raises here immediately. Due to the
well-known Lorentz-Fitzgerald contraction in the special relativity,
we would expect frame-dependence of $l$, which threatens its
``fundamentality''. There are two possibilities to remedy this
problem, and both are extensively studied in the Lorentz-violating
literatures; however, both are still controversial~\cite{m05,a08}.
The first is that, indeed, $l$ depends on the choice of frames, and
it transforms between different frames according to some laws (not
necessary to obey the conventional Lorentz transformation and the
symmetries of Poincar$\acute{\rm e}$ group). When the relevant
length scale is far larger than $l$, the transformation laws become
coincident with Lorentz transformation. The ``fundamentality'' of
$l$ relies on the existence of a preferred frame, which is often
chosen as the frame where the cosmic microwave background is
isotropic. Only in this preferred frame, observations agree on the
fundamental value of $l$~\cite{m05,a08}. The other possibility is
that observations within different frames agree on the same value of
$l$. A theoretical framework to be compatible with such an extra
physical constant is realized in doubly special relativity, firstly
proposed by Amelino-Camelia~\cite{a02,a02b,zsm11}. In this
framework, all observers situated in different inertial frames agree
on two fundamental quantities; the speed of light $c$ and a length
scale $l_{\rm DSR}$. Here we identify $l_{\rm DSR}$ as the $l$
discussed in the present paper. The ``fundamental'' meaning of $l$
is relativistic, even better achieved conceptually. Its position is
the same as the speed of light $c$ in the special relativity; hence,
$l$ and $c$ share the same conceptual significance in this approach.

From experimental aspects, Klinkhamer suggested four kinds of
experiments to determine different physical quantities, $l$ and
$\eta$ (or alternatively, $G$)~\cite{k10}. The traditional and
modern versions of the Cavendish experiment can measure the
Newtonian coupling constant $G \propto \eta^2l^2$. A complementary
way to disentangle $\eta$ and $l$ comes from the mentioned
Lorentz-violating research, where a separated determination of $l$,
or alternatively $E$ in Eq.~(\ref{le}), can be achieved from the
modified energy-momentum dispersion
relation~\cite{m05,a08,grb,xm09,sxm10,sm10,xsm10,fermi09,sm11prd,a02,a02b}.
The last two experiments considered in Ref.~\cite{k10} are {\it
gedanken experiments} from primordial gravitational waves and a
logarithmic correction of the entropy, which can probe an isolated
$l$, and, if proven correct and becomes practical, they can serve as
consistent checks to the above two practical means.

In summary, from Verlinde's conjecture that gravity is an emergent
force rather than a fundamental one, the Newtonian constant $G$ can
no longer function as a fundamental constant. The request also
merges from other ideas through contemplations on physical
foundations towards unification. Then it is natural to suggest a
fundamental length scale $l$ to replace the position of $G$. Such an
$l$ can be explained as the smallest length scale of quantum
space-time, and its value can be measured directly through searches
of Lorentz violation.

In principle, $l$ can be far away from the conventional Planck
length, $l_{P} \simeq 1.6 \times 10^{-35}$~m. The actual value of
$l$ has consequences in many physical aspects, especially in the
search of quantum gravity theories and ultimate understanding of the
structure of quantum space-time. The ``rescaling'' of $l$ from
$l_{P}$ can be understood through the detailed mechanism of
assigning bits of information onto holographic areas in the entropic
gravity approach. The rescaling of the fundamental length scale of
quantum space-time influences the conclusions drawn from
Lorentz-violating studies. The related experiments to disentangle
the involved physical quantities are also discussed, and the ones
from Lorentz-violating aspects are rather promising. It also can
provide new perspectives on the fundamental issue of basic
space-time units, hence the whole system of ``natural units''.

\begin{acknowledgments}
This work is partially supported by National Natural Science
Foundation of China (Nos.~11005018, 11021092, 10975003, 11035003).
\end{acknowledgments}

\end{document}